# Thermally driven elastic weakening of crystals

Jia Li, Yangxian Li, Xiao Yu, Wenjiang Ye

*School of Materials Science and Engineering, Hebei University of Technology, Tianjin 300130, China*

Chang Q Sun[*]

*School of Electrical and Electronic Engineering, Nanyang Technological University, Singapore 639798*

**Abstract**

An analytical solution has been developed clarifying that the thermally driven elastic softening of crystals can be directly related to the length and strength of the representative bonds of the crystal and to the response of the bonding identities to the change of temperature. Reproduction of the experimental results Ag, Au, MgO, Mg2SO4, Al2O3, and KCl derived mean atomic cohesive energy of the specimen may evidence the validity of the proposed approach without involving parameters using in classical thermodynamics.

---

[*] Fax: 65 6793 3318; Tel: 65 6790 4517; E-mail: Ecqsun@ntu.edu.sg

I    Introduction

Elastic modulus is one of the key elemental parameters in condensed matter physics and materials sciences, which is related to the performance of a material such as the elasticity, extensibility, acoustic transmission velocity, Debye temperature, specific heat capacity, and thermal conductivity of the specimen. As an elemental quantity of a solid, the elastic modulus and its temperature (T) dependence has been intensively investigated both experimentally and theoretically [1,2,3,4,5,6,7], because the temperature response of the modulus will result in the change of the above mentioned properties. Typically, the temperature dependence of the modulus follows Wachtman's relation [8],

$$Y = Y_0 - b_1 T \exp(-T_0 / T),$$

(1)

where $Y_0$ is the Young's modulus at T = 0 K. $b_1$ and $T_0$ are the arbitrary constants for data fitting. This model works quite well for the linear temperature dependence at high temperatures. However, at temperatures below the Debye temperature, $\theta_D$, the measured data manifest nonlinear temperature dependence. In order to improve the model, Andersen [9] derived an alternative form by introducing the Mie-Grüneisen equation into Wachtman's model,

$$Y = Y_0 - \frac{3R\gamma\delta T}{V_0} H(T/\theta_D),$$

$$H(T/\theta_D) = 3(T/\theta_D)^3 \int_0^{\theta_D/T} \frac{x^3 dx}{e^x - 1}$$

(2)

with $R$ being the ideal gas constant. $\gamma$ is Grüneisen parameter, $\delta$ is Anderson constant that is temperature independence. $V_0$ is the specific volume per mole of atoms at 0 K.

According to Anderson, the term $3RT\exp(-T_0/T)$ in eq (1) is virtually an empirical approximation of the inner energy of Debye approximation.

Recently, from the perspective of classical statistic thermodynamics, Garai et al. [10] derived a solution from the perspective of classical thermodynamics to fit the temperature dependence of the bulk modulus at 1 bar pressure,

$$B_T^0 = B_{T=0}^0 \times \exp\left[\int_{T=0}^{T} \alpha_V^0(T)\delta^0(T)dT\right]$$

(3)

where superscript 0 denotes quantities gained at 1 bar pressure. $B_{T=0}^0$ is the bulk modulus at T = 0 K, $\alpha_V^0$ is the volume coefficient of thermal expansion, and $\delta^0(T)$ is the Anderson-Grüneisen parameter.

Numerically, eqs (2) and (3) could reproduce the measurements at the entire temperature range of measurement despite the adjustable parameters such as the $\gamma$ and $\delta$ that are hardly known. On the other hand, in the contact mode of measurement such as nanoindentation, the compressive stress will enhance the measured elastic properties [11]. Unfortunately, the compressive stress effect was not considered in the currently available models [12]. In the present work, we present an analytical solution to include both the temperature and the pressure effect on the elastic modulus. The derived solution has enabled us to reproduce the measured temperature dependence of the elastic modulus of Ag, Au, MgO, $Mg_2SO_4$, $Al_2O_3$, and KCl crystals. On the other hand, the fitting to the experimental data leads to information of the mean atomic cohesive energy as output without other freely adjustable parameters being involved.

II   Principle: local bond average

The interatomic bonding is so important that it discriminates a solid from the isolated constituent atoms in physical properties [13]. It is therefore essential to correlate the detectable quantities to the bonding identities such as the order, nature, length and strength of the bonds involved. For a given specimen, no mater whether it is a crystal, non-crystal, or with defects or impurities, the nature and the total number of bonds do not change under the external stimulus, such as temperature or pressure, unless phase transition occurs. However, the length and strength of all the bonds involved will response to the stimulus. If the functional dependence of the detectable quantity on the bonding identities is known, one is able to predict the performance of the solid by focusing on the response of the length and energy of the representative bonds to the external stimulus. In practice, we may consider the average over all the bonds involved for the physical properties of the given specimen. This approach of local bond average (LBA) may represent the true situations of measurements or theoretical computations that collect statistic information from a large number of atoms of the given specimen. Furthermore, compared with the measurement and computation, the LBA could discriminate the behavior at different locations.

Considering the external stimulus of temperature and external pressure, the bond length $d(T,P)$ and the bond energy E(T, P) follow the relations:

$$d(T,P) = d(0,0) \times \left(1 + \int_0^T \alpha(t)dt\right) \times \left(1 + \int_0^P \beta(p)dp\right)$$

$$E(T,P) = E(0,0) - \int_0^T C(t)dt - \int_{V_0}^V pdV/V_0$$

$$= E(0,0) - \int_0^T C(t)dt + \int_0^P v(T,p)dp/V_0 - VP/V_0$$

(4)

where P is the external stress. The term $-\int_{V_0}^V pdV/V_0 = \int_0^P v(T,p)dp/V_0 - VP/V_0$ is the energy density stored to the crystal upon the sample is being pressed from $V_o$ to V [14]. The d(0,0) is the bond length at absolute 0 K without stress. α(t) is the temperature dependent thermal expansion coefficient. β(p) $= d\ln(d)/dp$ is the linear compressibility that is proportional to the inverse of elastic modulus. Evidence [15] shows that the β(p) remains constant at constant temperature within the elastic deformation regime and then the integration could be simplified as $\int_0^P \beta(p)dp \cong \beta P$. The $C(t)$ being the temperature-dependent specific heat per bond is assumed to follow Debye approximation, $C_v(T/\theta_D)$. The integral of $\int_0^T C(t)dt$ represents the increase of the inner energy due to the thermally-activated vibrations in all possible modes. The $v(T,P)$ is the average atomic volume. By assuming the atom as a hard sphere of diameter d, the T and P dependent volume and bond energy can be simplified as,

$$v(T,p)) \cong \pi d_0^3 (1+3\alpha T) \times (1+3\beta P)/6$$

$$E(T,P) \cong E(0,0) - \int_0^T C_v(T/\theta_D)dt + v_0 P(1+3\alpha T) \times \left(1 - v/v_0 + \frac{3\beta P}{2}\right)$$

because α and β are in the $10^{-6}$ order at room temperature. Therefore, the bonding energy contains three parts: (i) the bonding energy per bond at 0 K; (ii) the bond energy weakening due to thermal vibrations; (iii) the bond strengthening due to the storage of deformation energy caused by the compressive stress.

On the other hand, from the perspective of the LBA approach, the bulk modulus is proportional to the bond energy per unit volume [16],

$$B(T,P) = -\frac{1}{V}\left(\frac{\partial P}{\partial V}\right)_T = \left[-V \frac{\partial^2 U}{\partial V^2}\right]_T \propto \frac{E(T,P)}{d^3(T,P)}$$

(5)

Combining eqs (4) and (5) will lead to a immediate solution to the temperature and pressure dependence of the relative bulk modulus to the standard bulk value at T = 0 K and P = $P_0$ = 1 Pa,

$$\frac{B(T,P)}{B(0,0)} = \frac{E(0,0) - \int_0^T C_v(t/\theta_D)dt - v_0 \int_0^P \left[\left(1+\int_0^T \alpha(t)dt\right)\times\left(1+\int_0^P \beta(p)dp\right)\right]^3 dp}{E(0,0)\left[\left(1+\int_0^T \alpha(t)dt\right)\times\left(1+\int_0^P \beta(p)dp\right)\right]^3}$$

$$\cong \frac{E(0,0) - \int_0^T C_v(t/\theta_D)dt - v_0 p\left(1+3\int_0^T \alpha(t)dt\right)(1-v/v_0+\frac{3}{2}\beta p)}{E(0,0)\left(1+3\int_0^T \alpha(t)dt\right)\times(1+3\beta p)}$$

(6)

The specific volume at 0K per average atom, $v_0$, could be approximated by the ratio of average atomic mass $\bar{\mu}$ to the mass density $\rho$ of the specimen, i.e., $v_0 \cong \bar{\mu}/\rho$. The $E(0,0)$ is the only fitting parameter used to reproduce the temperature dependence of the elastic modulus. Most strikingly, neither Grüneisen nor Anderson parameter is needed in the current LBA model. However, the current model contains the contribution from thermal expansion and from the compressive stress presented in the measurement. Furthermore, the current LBA model specifies the relation of $E(0,0)^{-1} = 3R\gamma\delta/V_0$ relates to Anderson's model through the relation of

III  Results and Discussion

Using eq (6), we have analyzed the temperature dependence of the modulus of Ag, Au, MgO, $Mg_2SO_4$, $Al_2O_3$, and KCl under the isotropic standard pressure 1 bar (=$10^5$ Pa). The values of $\theta_D$, $\rho$, and $\beta$ [10] were used as the input parameters. The slight temperature dependence [16] of the input parameters may lead to a slight deviation of the derived E(0,0) from the true value. We first conducted the linear fit at higher temperature to estimate the E(0, 0) value. Further refinement of the E(0,0) can then be conducted by carefully fitting the measurement at the entire temperatures with the experimentally derived $\alpha(t)$ [10].

Figure 1 shows the reproduction of the measured temperature dependence of elastic modulus of different samples measured at 1 bar pressure. The associated input and output data are summarized in Table 1. Generally, at temperatures higher than the Debye temperature, the elastic modulus depends linearly on temperature; at sufficiently low temperatures the elastic modulus drops slowly with the increase of temperature because of the $T^4$ nature of the $\int_0^T C_v(t/\theta_D)dt$. It is therefore clear now that the $\theta_D$ determines the shoulder width and the E(0,0) determines the slope of the linear part of the B-T profile. The exceedingly good agreement evidences the validity of the current LBA approach. The fitting also provides a simple way for deriving information of the mean atomic cohesive energy. The accuracy of the E(0,0) depends merely on the accuracy of measurement in which the artifacts and impurities may be involved.

In order to illustrate the effect of pressure enhanced modulus, theoretical prediction has been made for Ag at 1 and 10 bar and shown in figure 1(d). Result demonstrates the pressure induced work hardening because of the storage of deformation energy. One can imagine that under pressure the bond length will be compressed and the deformation energy is stored into the bulk. Although no measurement data is immediately available, the predicted trend may improve our understanding of the pressure effect, which may inspire further modeling consideration of the pressure effect on the elastic modulus.

IV  Conclusion

An analytical solution has been developed for the temperature and pressure dependence of elastic modulus of crystals. It is shown that the response of the bonding identities to the external stimulus determines the thermally driven elastic softening. Reproduction of the measurement leads to information of the mean atomic cohesive energy, which is beyond the scope of approach is in terms of classical thermodynamics or the models with adjustable parameters such as the Gruneisen parameter or Anderson constant are involved in the model.

Table and Figure captions

Table 1 The input data including Debye temperature $\theta_D$, the T-dependent thermal coefficient $\alpha(t)$, the specific volume per "average" atom $v_0$, the compressibility $\beta$, and the optimized fitting parameter $E_B(0)$. In the bracket the corresponding temperature are shown.

|  | $\theta_D$ K | $v_0$ $10^{-29}$m$^3$/atom | $\rho$ g/cm$^3$ | $\beta$ $10^{-2}$GPa$^{-1}$ | $E_B(0)$ eV |
|---|---|---|---|---|---|
| Au | 170 | 1.69 | 19.3 | 0.44 | 1.64 |
| Ag | 160 | 1.71 | 10.5 | 0.63 | 1.24 |
| MgO | 885 | 0.96 | 3.5 | 0.71 | 1.29 |
| Al$_2$O$_3$ | 986 | 0.86 | 3.9 | 0.43 | 4.30 |
| Mg$_2$SO$_4$ | 711 | 0.84 | 3.1 | 0.91 | 2.80 |
| KCl | 214 | 3.24 | 1.9 | 7.14 | 0.57 |

Figure 1 Reproduction of the temperature dependence of the bulk modulus of various crystals (a-c, data are sourced from Ref 10) with derived information of atomic cohesive energy as listed in Table 1 and (d) prediction of the pressure and temperature effect on the modulus of Ag at 1 bar and 10 bar.

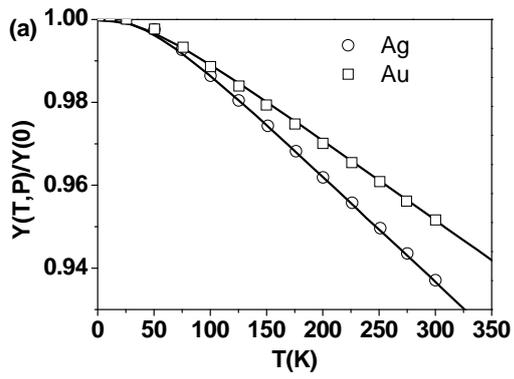
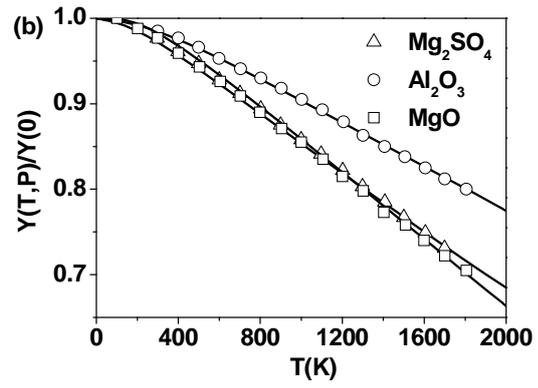
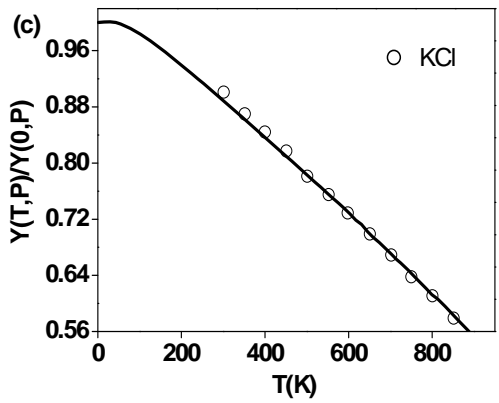
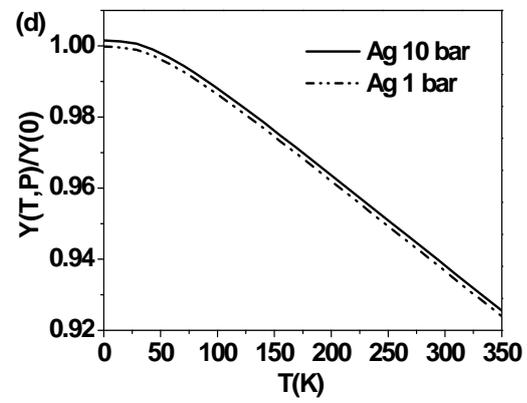